\documentclass[reprint, amsmath,amssymb, aps,pre,floatfix,]{revtex4-2}
\usepackage{graphicx}% Include figure files
\usepackage{dcolumn}% Align table columns on decimal point
\usepackage{bm}% bold math
\usepackage[position=t,singlelinecheck=off,justification=raggedright]{subfig}
\usepackage[]{natbib}
\usepackage[colorlinks,citecolor=blue]{hyperref}

\begin{document}

\title{Spin-split flat bands at the band edge and two-dimensional hole gases towards quantum Hall effect in altermagnetic CoF$_2$}

\author{Bo-Wen Yu}
\affiliation{Beijing National Laboratory for Condensed Matter Physics, Institute of Physics, Chinese Academy of Sciences, Beijing 100190, China}
\affiliation{School of Physical Sciences, University of Chinese Academy of Sciences, Beijing 100049, China}
\author{Bang-Gui Liu}\email{Email: bgliu@iphy.ac.cn}
\affiliation{Beijing National Laboratory for Condensed Matter Physics, Institute of Physics, Chinese Academy of Sciences, Beijing 100190, China}
\affiliation{School of Physical Sciences, University of Chinese Academy of Sciences, Beijing 100049, China}


\date{\today}

\begin{abstract}
Altermagnetic phase is recently found as a new magnetic phase in addition to the conventional collinear spin orders, and great efforts have been made to explore novel effects and potential applications in  such materials. Here, we show that there are robust altermagnetic spin-split flat bands near the valence band edge in rutile CoF$_2$  through first-principles investigation. It is uncovered that the magnetic moments can remain in the z axis because of the magnetocrystalline energy due to the spin-orbits coupling and the spin orientation can be made more stable by magnetic field applied in the xy plane. We describe the spin-dependent band structure (including the flat bands) near the Fermi level by a spin-resolved effective low-energy model, and reveal that they can host spin-dependent two-dimensional hole gases. 
Importantly, we find spin-dependent quantum Hall effects in the two-dimensional hole gases by applying the magnetic field in the xy plane, and then explore the dependence of Hall conductivity and Hall resistance on the Fermi level and the magnetic field (both magnitude and direction) and related longitudinal carrier transport properties.
\end{abstract}

\maketitle

\section{Introduction}

Conventionally, collinear magnetic orders are divided into ferromagnetic, ferrimagnetic, and antiferromagnetic phases. Recently a new phase, namely altermagnet , is introduced to describe a special magnetic state, which is seen as a breakthrough in the field of fundamental magnetism during the past few years\cite{PRXam1,PRXam2,LJW}. This new phase of magnetism can be understood in terms of spin space groups \cite{P_SpinGroup,ACSA_SpinGroup2,PRX_SpinSpaceGroups,PRX_SpinSpaceGroups2,PRX_SpinSpaceGroups3}. Many conventional antiferromagnets can be considered candidates for altermagnetic materials with the help of systematic analysis\cite{PRX_AM_RV}. So far, some systems have been experimentally confirmed to be altermagnetic materials, such as Mn$_5$Si$
_3$ \cite{Mn5Si3-LJW,NC_anomalousHall_Mn5Si3}, CrSb \cite{NC_AM_CrSb,PRM_AM_CrSb,CrSbadd1}, Fe$_2$Se$_2$O \cite{nanole_AM_Fe2Se2O}, MnTe\cite{Nat_AM_MnTe_LKST,PRL_AM_MnTe_LKST2,PRB_AM_MnTe,MnTenature}, and MnTe$_2$\cite{Nat_AM_MnTe2}, and others such as RuO$_2$ are still under debate for possible  altermagnetism\cite{PRL_RuO2_1,PRL_RuO2_NonmagGrdState,PRL_RuO2_SpinHallEffect,PRB_RuO2_DFTU,arxiv_LJW_RuO2, Mokrousov_RuO2_CoF2,PRL_RuO2_8Absence}. In addition, more materials are predicted theoretically to be altermagnetic\cite{MTP_HighThroughput,PRX_Octupoles_MnF2,2DM_AM_RuF4}. 
It has been established that there exists nonrelativistic spin-splitting and spin-momentum coupling in altermagnetic materials \cite{PRL_NonrelSpinSplit_1,PRX_NonrelSpinSplit_2,NC_SpinCurrent1, NE_RuO2_6,PRL_RuO2_SpinTorque}. These  features can be used to realize electric reversal of Neel vectors\cite{Mn5Si3-LJW}, spin-polarized currents  \cite{NC_SpinCurrent2,PRB_Perovskite_SpinSplit,SA_RuO2_1,PRL_NonrelSpinSplit_3}, spin-splittering effect\cite{PRL_NonrelSpinSplit_1,spinfiltering}, and unconventional superconductivity\cite{SupC1,SupC2,SupC3,SupC4,SupC5,SupC6,SupC7,SupC8,SupC9,SupC10}. 
Further research and exploration are highly desirable in order to seek more effects and potential applications for electronics, spintronics and multiferroics.

Rutile CoF$_2$ is famous as an antiferromagnetic wide band gap semiconductor, and its fundamental properties have been studied theoretically and experimentally\cite{JACS_CoF2Structure,PR_CoF2_1,PR_CoF2_2_Magnetic,PR_CoF2_3,JPSJ_CoF2_4, JAP_CoF2_5,NP_CoF2_7_Spintronics,PRB_CoF2_6,PRB_CoF2_mag}. Its altermagnetic spin splitting of electronic bands have been shown through ab-initio study\cite{MTP_HighThroughput}, and some effort has been made to explore its potential spintronic and spin-filtering applications\cite{NP_CoF2_7_Spintronics,spinfiltering}. 
Here, by performing systematic first-principles investigation of the magnetic properties and electronic structure of rutile CoF$_2$, we shall show that there are robust spin-split flat bands near the valence band edge and these unique band structure can host spin-resolved two-dimensional hole gases. Furthermore, we find spin-dependent quantum Hall effects in these special hole gases with magnetic field applied in the xy plane and explore the dependence of the Hall conductivity and Hall resistance on the Fermi level and applied magnetic field (both magnitude and direction). Computational methods and more detailed results will be presented in the following.

\section{Methodology}

The first-principles calculations are performed with the projector-augmented wave (PAW) method within the density functional theory\cite{PAW}, as implemented in the Vienna Ab-initio simulation package software (VASP) \cite{VASP}. The generalized gradient approximation (GGA) revised for solids by Perdew, Burke, and Ernzerhof (PBEsol)\cite{PBE,PBEsol} is used as the exchange-correlation functionals. The  self-consistent calculations are carried out with a $\Gamma$-centered ($12\times 12\times 19$) Monkhorst-Pack grid\cite{MPgrid}. The kinetic energy cutoff of the plane wave is set to 450 eV. The convergence criteria of the total energy and force are set to 10$^{-7}$ eV and 0.001 eV/\AA{}. The spin-orbit coupling (SOC) is taken into account in the calculation of electronic band structures and magnetocrystalline anisotrpy energy. The electronic correlation  is considered to improve energy band description \cite{HubbardU}.

\begin{figure*}%[h]
 \includegraphics[width=0.7\textwidth]{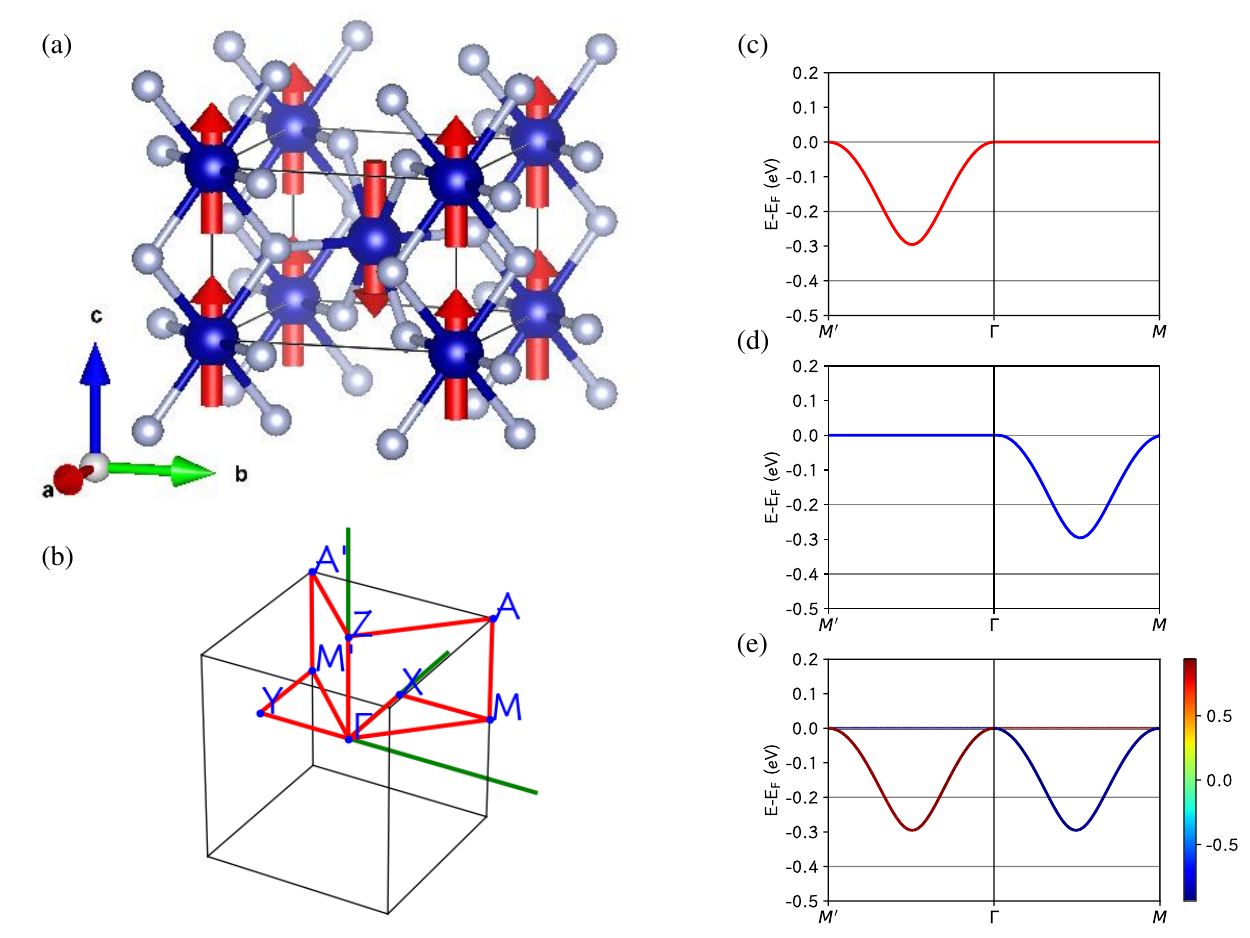}
\caption{ (a) The structures of rutile CoF$_2$ in the ground state phase. The Co atoms are located at the center and the corners of the unit cell, and the arrows indicate spin directions.  There are two sublattices in the unit cell. (b) The corresponding Brillouin zone of the rutile CoF$_2$. (c-e) The energy band structures along M-$\Gamma$-M' of the rutile CoF$_2$ without SOC (c, spin-up; d, spin-down) and  with SOC (e), with the complete band strctures presented in Fig. S1 (Supplementary Materials). The color bar in (e) describes the electronic spin polarization, with ±1 corresponding to the 100\% spin polarizations of spin-up or spin-down. There exists huge nonrelativistic spin splitting along M-$\Gamma$-M' in the Brillouin zone. The spin-up  and spin-down flat bands  along M-$\Gamma$-M' make the valence band maxima. }\label{fig1}
\end{figure*}

\section{Result and discussion}

% \subsection{Structures and altermagnetism}

The crystal structure and basic properties of the bulk CoF$_2$ have been studied theoretically and experimentally \cite{JACS_CoF2Structure,PR_CoF2_1,PR_CoF2_2_Magnetic,PR_CoF2_3,JPSJ_CoF2_4,JAP_CoF2_5, NP_CoF2_7_Spintronics,PRB_CoF2_6,PRB_CoF2_mag,MTP_HighThroughput}. 
Theoretical investigation and X-ray diffraction experiments show that bulk CoF$_2$ assumes a rutile structure\cite{JACS_CoF2Structure}, as described in Fig. \ref{fig1}(a). The experimental lattice constants are $a_e = 4.695$ \AA{} and $c_e = 3.180$  \AA{} \cite{JACS_CoF2Structure,PR_CoF2_1}. There are 6 atoms in the unit cell of rutile CoF$_2$. The two Co atoms are located at (0,0,0) and (0.5,0.5,0.5), and experimentally measured magnetic moments are collinear 2.21 $\mu_B$ and -2.21 $\mu_B$, with magnetic easy-axis along the z axis, so that the total magnetization of rutile CoF$_2$ is eqivalent to zero\cite{PRB_CoF2_mag}. The two Co atoms and their neighboring F atoms can be divided into two sublattices. In terms of the tetragonal symmetry, the two sublattices can be transformed into each other through a twofold spin-space rotation with a fourfold real-space rotation and a translation operation\cite{PRX_AM_RV,P_SpinGroup,ACSA_SpinGroup2}. The corresponding Brillouin zone is also presented in Fig. \ref{fig1}(b).

Our theoretically optimized equilibrium lattice constants are $a_t = 4.646$ \AA{} and $c_t = 3.157$ \AA{}, and the total magnetic moment of the unit cell is zero.  The Co atom has 9 electrons in the 3d and 4s orbitals, and there are 4 double-occupied d electrons and 3 single-occupied d electrons per Co atom after bonding with the 2+ valence in the rutile CoF$_2$ \cite{PRB_CoF2_mag}. The 3 single-occupied electrons contribute to the magnetic  moment, and then the Co ion has spin $S=\frac 32 \hbar$. 
In order to confirm the magnetic easy axis, we calculate the total energies along different axises. The calculated results show that the lowest total energy corresponds the ground state with the collinear magnetic moments orienting in  the [001] axis,  and the [100] and [110] axises are higher than the ground state by 0.2meV and 1.2meV per unit cell, respectively.

Several methods including HSE06\cite{HSE06}, PBE+U and PBEsol+U\cite{HubbardU,PBEsol} are used for comprehensive evaluation of the band structure. The results with different $U$ values and HSE06 method are presented in  Fig. S1 (Supplementary Materials). Considering the gap calculated by HSE06 is generally larger than the real situation, we believe that $U=3$eV should be the most appropriate value for the band description. The spin-resolved electronic band structures of the optimized ground-state structure are presented in  Fig. S2 (a,b), being consistent with previous results\cite{MTP_HighThroughput}, and part of them along M-$\Gamma$-M' are plotted in Fig. 1(c,d). The semiconductor gap of the rutile CoF$_2$ is 2.5 eV, indicating that it is a wide-bandgap semiconductor. The spin-resolved band structure without SOC show that the CBM is at point A in the Brillouin zone for both spin up and down and the VBM is at the line $\Gamma$-M (0.5,0.5,0) for spin up and the line $\Gamma$-M' (-0.5,0.5,0) for spin down. At the VBM, the spin-up flat band is along the line $\Gamma$-M, while the spin-down flat band is along the line M'-$\Gamma$. It is very interesting that the flat bands are at the VBM and make the valence band edge!
Careful comparison of the band structures without and with SOC shows that this inter-spin band-edge splitting does not come from SOC. The band structure with SOC is also presented in Fig. S2 (c) and Fig. 1(e). It is clear that the flat bands survive when the effect of SOC is taken into account.
The flat bands at the VBM implies that doped holes have a huge effective mass and can hardly move in that direction, and on the other hand they can be considered to be degenerate and used to make two-dimensional hole gases. 

\begin{figure}%[h]
\includegraphics[width=0.95\columnwidth]{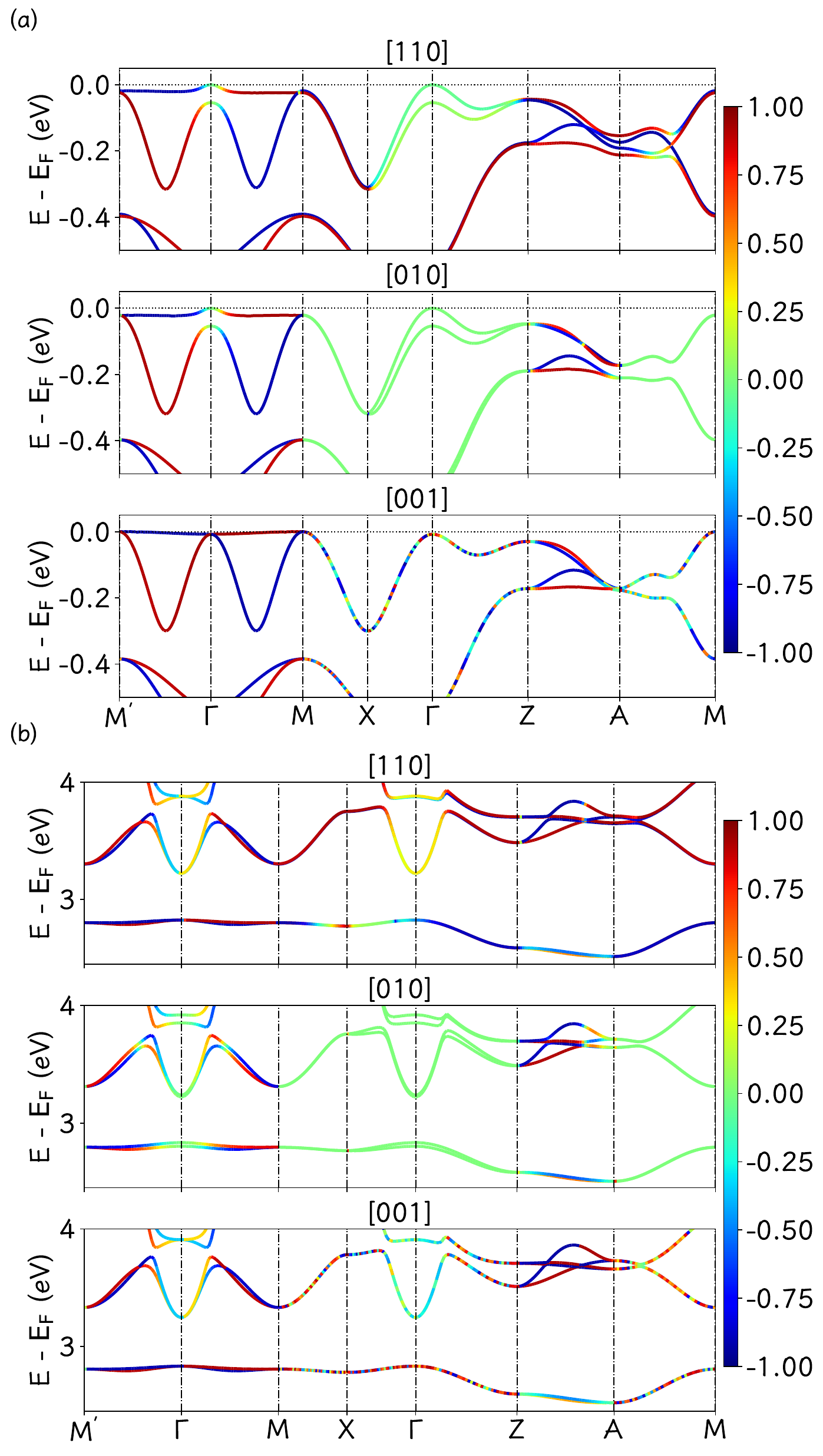}
\caption{ The electronic band structures of the CoF$_2$ near the valence band maxima (a) and the conduction band minima (b), with the magnetic moment orienting in the [100], [110], and [001] axises. The color bar describes the electronic spin polarization, with ±1 corresponding to the 100\% spin polarizations of spin-up or spin-down. The flat bands at the VBM and the nonrelativstic inter-spin band-edge splitting are both maintained in the ground-state phase with spins orienting along [001] axis.}\label{fig2}
\end{figure}

We calculate the electronic structures of the CoF$_2$ with the magnetic moment orienting in the [100], [110], and [001] directions, and present the SOC band structures near the valence and conduction band edges in  Fig. \ref{fig2}. It is clear that the flat bands remain at the valence band edge (VBM) when the magnetic moments are in the [001] direction. When the magnetic moments are made different from the [001] direction, there will be no flat bands at the $\Gamma$ point and there will be some flat bands below the VBM (away from the $\Gamma$ point). Therefore the electronic properties of the energy band edges can be controlled by changing the direction of the magnetic moments.

In order to analyze the relationship between the spin amd orbitals of different Co atoms, we present the band structures with atom and orbital projections for the two Co atoms in Fig. S3. There are only 7 electrons occuping the five 3d orbitals for the Co ion. It is clear that the lowest conduction bands originate from Co d$_{x^2-y^2}$ orbitals, but there is no contribution from the Co d$_{x^2-y^2}$ orbitals in the valence bands between -2 eV and the Ferni level. The flat bands at the VBM can be attributed to the linear combination of d$_{xz}$ and d$_{yz}$. The next two valence bands can be attributed to a linear combination of d$_{xy}$ and d$_{z^2}$. The lower valenve bands originate from these four orbitals of the two Co ions located at (0.5,0.5,0.5) and (0,0,0). The inter-spin band splitting in the Brillouin zone is connected with the two Co ions with different spin orientations in the real space.

% \subsection{Carrier charge and spin currents through the electric and magnetic field}

\begin{figure}%[h]
\includegraphics[width=\columnwidth]{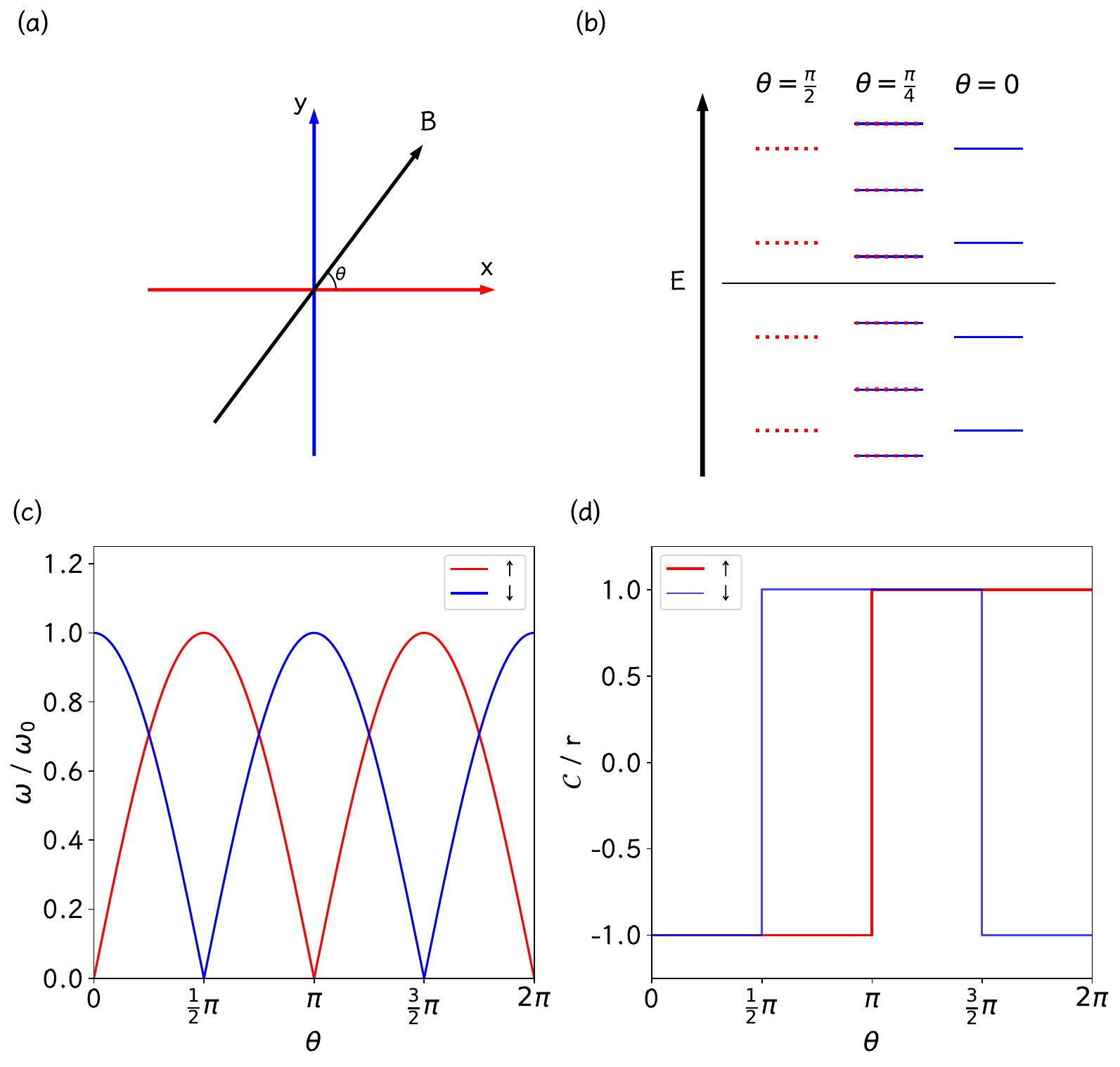}
\caption{ (a) The direction ($\theta$) of external magnetic field in the xy plane formed with the x and y axises. (b) The Landau levels for spin-up (red) and spin-down (blue) electrons for $\theta=\frac{\pi}{2}$, $\frac{\pi}{4}$ and $0$.  (c) The spin-dependent parameter $\omega$ as  functions of $\theta$.  (d) The spin-dependent Chern numbers per occopied band as functions of $\theta$. }\label{fig3}
\end{figure}

Special band structure causes electrons to exhibit interesting properties, especially in the external electric and magnetic field. The flat bands near the VBM can make the doped holes act as two-dimensional hole gases. For convenience, we reset the definition of the axises. The z axis is still along the c direction of the lattice, and the x and y axises are defined to be along the two diagonals of the xy plane, as shown in Fig. \ref{fig3} (a). From the valence band structure near the VBM, we can construct a spin-resolved effective low-energy model. The Hamiltonians for spin up and spin down can be written as
\begin{equation}
\begin{array}{l}
\displaystyle H_{\uparrow}=\frac{(\hat{p}_x)^2}{2M_0}+\frac{(\hat{p}_z)^2}{2M_z}, ~~
\displaystyle H_{\downarrow}=\frac{(\hat{p}_y)^2}{2M_0}+\frac{(\hat{p}_z)^2}{2M_z},
\end{array}
\end{equation}
where $M_0$ is the effective mass of the hole along the x axis for spin up (y axis for spin down) and $M_z$ is the effective mass along the z axis. The $M_0$ and $M_z$ can be obtained by fitting the curves of the band structure near the valence band edge. As a result, $M_0$ is 0.1603 $m_e$ and $M_z$ is 0.2461 $m_e$, where $m_e$ is the free-electron mass. The different spins are studied independently because the inter-spin intersction can be neglected. This Hamiltonian set describes two independent 2D holes gases in the two different planes. 

We apply a magnetic field in the xy plane, $\bm{B}=(B_x,B_y,0)$. Because it is perpendicular to the magnetic  moments, there will be no spin flop\cite{sflop1,sflop2,sflop3,sflop4,sflop5}. The gauge of vector potential $\bm{A}=(A_x,A_y,A_z)$ can be made depend on the spin orietation, and thus we have $\bm{A}_{\uparrow}=(0,zB_x,-xB_y)$ for spin up and $\bm{A}_{\downarrow}=(zB_y,0,yB_x)$ for spin down. Consequently, the Hamiltonians can be expressed as
\begin{equation}
\left\{ \begin{array}{l}
\displaystyle H_{\uparrow}=\frac{\hat{p}_x^2}{2M_0}+\frac{q^2 B_y^2}{2M_z}(\hat{x}-\frac{\hat{p}_z}{qB_y})^2\\
\displaystyle H_{\downarrow}=\frac{\hat{p}_y^2}{2M_0}+\frac{q^2 B_x^2}{2M_z}(\hat{y}-\frac{\hat{p}_z}{qB_x})^2.
\end{array}\right.
\end{equation}
The magnetic-field-induced magnetic moments must be tiny and the Zeeman terms can be neglected because the magnetocrystalline energy is large.
It is clear that the Hamiltonians are like the oscillator Hamiltonians and the essential parameters of the Landau levels are $\omega_{\uparrow,\downarrow}=\frac{|qB_{y,x}|}{\sqrt{M_z M_0}}$. The energy levels for spin up and spin down are $E_{m,\uparrow}=\hbar \omega_{\uparrow}(m+\frac{1}{2})$ and $E_{m,\downarrow}=\hbar \omega_{\downarrow}(m+\frac{1}{2})$, where $m$ describes the Landau levels. It is indicated that the energy levels for different spins are dependent on the direction ($\theta$) and strength $B$ of the magnetic field. 

We plot  the Landau levels for three different $\theta$ values in Fig. \ref{fig3} (b). For $\theta=\frac{\pi}{4}$, the magnetic field applies to both spins equally, and their Landau levels are the same since the Zeeman effect is neglected. For $\theta=0$ or $\theta=\frac{\pi}{2}$, the magnetic field is perpendicular to either of the planes of the 2D hole gases and there will be only one set of Landau levesls. The energy difference of Landau levels depends on the parameters $\omega_{\uparrow}=\omega_0|\sin \theta|$ and $\omega_{\downarrow}=\omega_0|\cos \theta|$, where $\omega_0=\frac{|qB|}{\sqrt{M_z M_0}}$. The trend of $\omega_{\uparrow}$ and $\omega_{\downarrow}$changing with  $\theta$ is shown in Fig. \ref{fig3} (c). 
When the first $r_{\uparrow}+r_{\downarrow}$ Landau levels are occupied, the Berry curvatures are $\Omega_{xz,\uparrow}(k_x,k_z)=-\frac{r_{\uparrow}}{|qB|\sin{\theta}}$ for spin up and $\Omega_{yz,\downarrow}(k_y,k_z)=-\frac{r_{\downarrow}}{|qB|\cos{\theta}}$ for spin down and the corresponding Chern numbers are \cite{PRB_2dBerryChern}
\begin{equation}
C_{\uparrow}=-r_{\uparrow}{\bf{sign}} (\sin{\theta}), ~~~
C_{\downarrow} =-r_{\downarrow}{\bf{sign}} (\cos{\theta}).
\end{equation}
The Chern numbers per occupied bands for the two spin as functions of $\theta$ are shown in Fig. \ref{fig3} (d).

% $M_0=0.1603 m_e$ and $M_z=0.2461 m_e$
\begin{figure}[tb]
\includegraphics[width=\columnwidth]{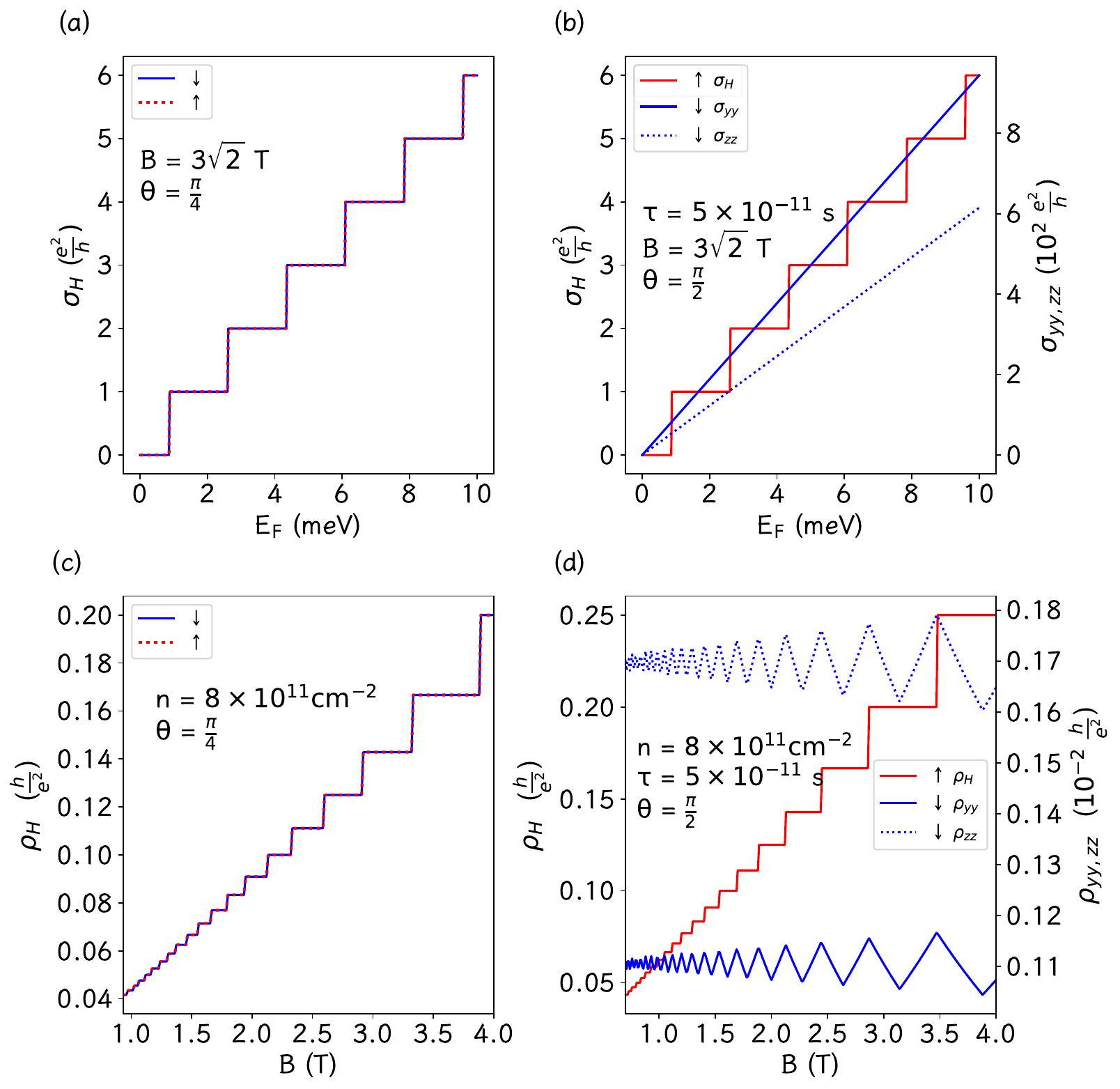}
\caption{ (a) The spin-resolved Hall conductivity ($\sigma_H$)  for $\theta=\frac{\pi}{4}$ as functions of the Fermi level $E_F$.  (b) The Hall conductivity of spin up ($\sigma_H$) and the longitudinal conductivity of spin down ($\sigma_{yy}$, $\sigma_{zz}$) for $\theta=\frac{\pi}{2}$ as functions of $E_F$.  (c) The spin-resolved Hall resistances ($\rho_H$) for $\theta=\frac{\pi}{4}$ as functions of $B$. (d) The Hall resistance of spin up ($\rho_H$) and the longitudinal resistances of spin down ($\rho_{yy}$, $\rho_{zz}$) for $\theta=\frac{\pi}{2}$ as functions of $B$. }\label{fig4}
\end{figure}

On the other hand, we discuss electronic transport without Landau levels. We apply a weak electric field $\bm{E}=(E_x,E_y,E_z)$ and assume that concentration of carriers $n$ is small enough, in order to keep the band structures unchanged. According to the the relaxation time approximation, the Boltzmann equation of the distribution function $f$ can be expressed as $\frac{f_k^0-f_k}{\tau}=\frac{\partial f_k}{\partial t}=\frac{\partial f_k}{\partial E_k}\frac{\partial E_k}{\partial t}=\frac{\partial f_k}{\partial E_k}\bm{v}_k\cdot \bm{E}q$, where $\tau$ is the electron relaxation time and $\bm{v}_k = \frac{d\omega}{d\bm{k}} = \frac{1}{\hbar}\frac{dE(k)}{d\bm{k}}$. The longitudinal conductivity components can be written as\cite{PRB_2dLongConductivity}
\begin{equation}
\left\{ \begin{array}{l}
\displaystyle         \sigma^{\uparrow}_{xx}=\frac{q^2\tau n}{M_0}, ~~ \sigma^{\uparrow}_{yy}=0, ~~
        \sigma^{\uparrow}_{zz}=\frac{q^2\tau n}{M_z} \\
\displaystyle         \sigma^{\downarrow}_{xx}=0 ,~~  \sigma^{\downarrow}_{yy}=\frac{q^2\tau n}{M_0},~~
        \sigma^{\downarrow}_{zz}=\frac{q^2\tau n}{M_z}.
\end{array}\right.
\end{equation}
The longitudinal conductivity is highly related to the spin orientation and electric field direction. 
If the electric field along the z axis, the longitudinal conductivity is the same for the both spins, $j^{\uparrow} =  |E|\sigma^{\uparrow}_{zz}$ and $j^{\downarrow} =  |E|\sigma^{\downarrow}_{zz}$, where $|E|$ is the magnitude of the electric field, and as a result, there is no spin-polarized current. 
If the electric field is in the xy plane, the longitudinal current is dependent on spin and the spin-resolved electronic current density can be written as
$j^{\uparrow} =  |E|\cos^2{\theta_1} \sigma^{\uparrow}_{xx}$ and $j^{\downarrow} =  |E|\sin^2{\theta_1} \sigma^{\downarrow}_{yy}$,
where $\theta_1$ is the angle of the electric field with respect to the x axis. The polarization of the spin-resolved currents is equivalent to $\frac{j^{\uparrow}-j^{\downarrow}}{j^{\uparrow}+j^{\downarrow}}=\cos{(2\theta_1)}$, which indicates that the polarization can be manipulated by changing the direction of the electric field in the xy plane.

It is interesting to study the corresponding Hall conductivity $\sigma_H$ and Hall resistance $\rho_H$ as functions of the Fermi level $E_F$ and magnetic field $B$ for given angle $\theta$ and concentration of carriers $n$. Here we use $n=8\times \mathrm{10^{11} cm^{-2}}$. For $\theta=\pi/4$, we have $\omega_{\uparrow}=\omega_{\downarrow}$ and $C_{\uparrow}=C_{\downarrow}$, and then both $\sigma_H$ and $\rho_H$ are the same for the spin up (in the xz plane) and spin down (in the yz plane). The calculated $\sigma_H$ and $\rho_H$ results are presented in Fig. \ref{fig4} (a,c). For $\theta=\pi/2$, we have Landau energy levels for spin up (the 2D hole gas in the xz plane) only, and there is no Landau levels for the 2D hole gas of spin down in the yz plane. The $\sigma_H$ and $\rho_H$ for spin up are presented in Fig. \ref{fig4} (b,d). The corresponding longitudinal conductivity ($\sigma^{\downarrow}_{yy}$ and $\sigma^{\downarrow}_{zz}$) and resistance  ($\rho^{\downarrow}_{yy}$ and $\rho^{\downarrow}_{zz}$) are also indicated in Fig. \ref{fig4} (b,d), with the relaxation time $\tau= 5 \times \mathrm{10^{-11} s}$. For $\theta=0$, Landau levels can be formed only for spin down in the yz plane, and the calculated results are similar to those of  $\theta=\pi/2$. 

\begin{figure}[tb]
\includegraphics[width=\columnwidth]{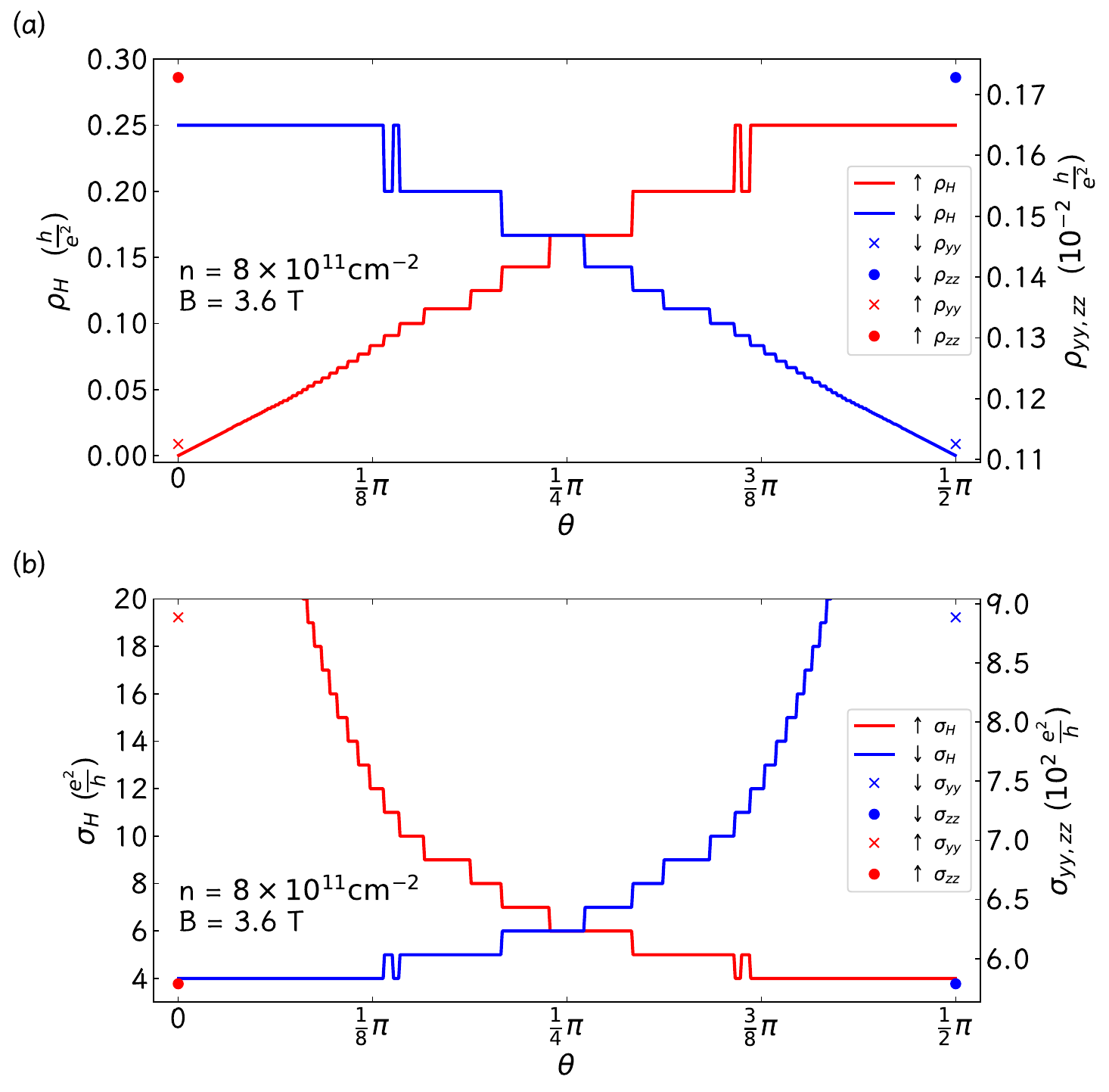}
\caption{ (a) The spin-dependent Hall resistance ($\rho_H$) (a) and  Hall conductivity ($\sigma_H$) (b) as functions of $\theta$. The spin up (down) is in red (blue). The spin-dependent longitudinal resistance and conductivity ($\theta=0,\pi/2$) are also presented. }\label{fig5}
\end{figure}

Furthermore, we present the $\theta$ dependence of the the Hall resistance and Hall conductivity in Fig. \ref{fig5}, with the longitudinal components indicated for comparison. It is clear that we have  spin degeneracy near the point $\theta=\pi/4$, and there are large difference between the two spin channels when $\theta$ is away from $\pi/4$. It is interesting that the Hall conductivity (resistance) of spin down in the yz plane increases (decreases) with $\theta$, except for a narrow `abnormal' region near $\theta=\pi/8$. Actually, there is a symmetry  in the curves between spin up (red) and spin down (blue) for $\theta\rightarrow\pi/2-\theta$. The narrow valleys and peaks near $\theta=\pi/8$ and $\theta=3\pi/8$ are formed because of the $\theta$-dependent ballance of the carriers with two spins in occupying the Landau levels, as shown in Fig. S4.

\section{Conclusion}

In summary, we have shown that there are robust spin-split flat bands near the valence band edge in the altermagnetic CoF$_2$  through first-principles investigation. The magnetic moments can remain in the z axis because of the magnetocrystalline energy due to the effect of SOC and their orientation can be made more stable by magnetic field applied in the xy plane. Furthermore, it has been established that the spin-dependent band structure (including the flat bands) near the Fermi level can be well described by a spin-resolved effective low-energy model, and these unique band structure can host spin-dependent two-dimensional hole gases in the xz and yz planes. 
Most importantly, after investigating the transport properties of the two two-dimensional hole gases under external electric and magnetic fields, we find spin-dependent quantum Hall effects by applying the magnetic field in the xy plane, and then explore the dependence of Hall conductivity and Hall resistance on the Fermi level and the magnetic field (both magnitude and direction) and related longitudinal carrier transport properties. It is believed that more interesting effects and phenomena will be achieved in such flat-band-induced two-dimensional carrier gases in three-dimensional bulk semiconductor materials.

\begin{acknowledgments}
This work is supported by the Strategic Priority Research Program of the Chinese Academy of Sciences (Grant No. XDB33020100) and the Nature Science Foundation of China (Grant No.11974393).  All the numerical calculations were performed in the Milky Way \#2 Supercomputer system at the National Supercomputer Center of Guangzhou, Guangzhou, China.
\end{acknowledgments}

% \bibliographystyle{a1.bst}
% \bibliographystyle{apsrev4-1}
% \usepackage[colorlinks=false]{hyperref}

\bibliographystyle{apsrev4-2}

\bibliography{ybw2}
% \printbibliography

\end{document}